\documentclass[%
 reprint,
 amsmath,amssymb,
 aps,
]{revtex4-1}

\usepackage{graphicx}
\usepackage{dcolumn}
\usepackage{bm}

\begin{document}

\title[Minimal length, maximal momentum and Hilbert space representation of quantum mechanics]
{Minimal length, maximal momentum and Hilbert space representation
of quantum mechanics}

\author{{ Kourosh Nozari \footnote{knozari@umz.ac.ir}}\quad and\quad  { Amir Etemadi  \footnote{a.etemadi@stu.umz.ac.ir}}}

\address{Department of Physics, Faculty of Basic Sciences,\\ University of Mazandaran, P. O. Box 47416-95447,\\ Babolsar, IRAN}

\begin{abstract}
Kempf {\it et al.} in Ref.\cite{1} have formulated a Hilbert space
representation of quantum mechanics with a minimal measurable
length. Recently it has been revealed, in the context of doubly
special relativity, that a test particles' momentum cannot be
arbitrarily imprecise and therefore there is an upper bound for
momentum fluctuations. Taking this achievement into account, we
generalize the seminal work of Kempf {\it et al.} to the case that
there is also a maximal particles' momentum. Existence of an upper
bound for the test particles' momentum provides several novel and
interesting features, some of which are studied in this paper.
\end{abstract}

\maketitle

\section{Introduction}
It is now a well-known issue that gravity induces uncertainty.
Incorporation of gravity in quantum field theory leads naturally
to an effective cutoff (a minimal measurable length) in the
ultraviolet regime. In fact, the high energies used to probe small
distances significantly disturb the spacetime structure by their
powerful gravitational effects. Some approaches to quantum gravity
such as string theory \cite{2}-\cite{8}, loop quantum gravity
\cite{9} and quantum geometry \cite{10} all indicate the existence
of a minimal measurable length of the order of the Planck length,
$l_{pl}\sim 10^{-35}m$ (see also \cite{11}-\cite{13}). Moreover,
some Gedanken experiments in the spirit of black hole physics have
also supported the idea of existence of a minimal measurable
length \cite{14}. So, existence of a minimal observable length is
a common feature of all promising quantum gravity candidates. The
existence of a minimal measurable length modifies the Heisenberg
uncertainty principle (HUP) to the so called Generalized
(Gravitational) uncertainty principle (GUP). In HUP framework
there is essentially no restriction on the measurement precision
of the particles' position so that $\Delta x_{0}$ as the minimal
position uncertainty could be made arbitrarily small toward zero.
But this is not essentially the case in the GUP framework due to
existence of a minimal uncertainty in position measurement. The
presence of the minimal observable length also modifies the
Hamiltonian of physical systems leading to Planck scale
modification of the energy spectrum of quantum systems. This issue
stimulated a lot of research programs in recent years, some of
which are addressed in Refs. \cite{15}-\cite{31}. One can adopt
the concept of the existence of a minimal observable length as a
nonzero minimal uncertainty $\Delta x_{0}$ in position
measurements. This feature would be a route to the noncommutative
structure of spacetime at the Planck scale and makes spacetime
manifold to have a foam-like structure at this scale. Based on
these arguments, one cannot probe distances smaller than the
Planck length in a finite time. In an elegant paper, Kempf {\it et
al.} have formulated the Hilbert space representation of quantum
mechanics in the presence of a minimal measurable length \cite{1}.
This work has been the basis of a large number of research
programs in recent years.

On the other hand, Doubly Special Relativity theories (for review
see for instance \cite{32}) suggest that existence of a minimal
measurable length would restrict a test particles' momentum to
take any arbitrary values leading to an upper bound, $P_{{{\it
max}}}$, on this momentum. This means that there is a maximal
particles' momentum due to fundamental structure of spacetime at
the Planck scale \cite{33}-\cite{36}. Here we are going to
generalize the seminal work of Kempf {\it et al.} to the case that
the existence of a maximal particles' momentum is considered too.
This extra ingredient brings a lot of new features to the Hilbert
space representation of quantum mechanics at the Planck scale. We
note that a more general case includes also a nonzero,
\emph{minimal} uncertainty in momentum measurement as well as
position. However, this general case is far more difficult to
handle since neither a position nor a momentum space
representation is available (see Ref.\cite{1} for more details).
Here we consider the case that there is just a minimal uncertainty
in position and particles' momentum is restricted also to the
upper bound $P_{{{\it max}}}$. By allowing the minimal uncertainty
in momentum to vanish, the Heisenberg algebra can be studied in
momentum space easily. In this manner, we can explore the quantum
physical implications and Hilbert space representation in the
presence of a minimal measurable length and a maximal particles'
momentum. We compare our results with Kempf {\it et al.} work
\cite{1} in each step.

\section{A brief about GUP}

\subsection{GUP with a minimal length}

In ordinary quantum mechanics, the standard Heisenberg Uncertainty
Principle (HUP) is given by
\begin{equation}
\Delta x \Delta p \geq \frac{\hbar}{2}\,.
\end{equation}
There is no trace of gravity in this relation. Today we know that
HUP breaks down for energies close to the Planck scale where the
corresponding Schwarzschild radius becomes comparable with the
Compton wavelength and both becoming approximately equal to the
Planck length. By taking into account the gravitational effect,
emergence of a minimal measurable distance is inevitable. This is
encoded in the following Generalized(Gravitational) Uncertainty
Principle (GUP) \cite{29}
\begin{equation}
\Delta x \Delta p \ \geq\ \frac{\hbar}{2}\ + \beta_{0} l_{pl}^2
\frac{(\Delta p)^2} {\hbar}\,.
\end{equation}
The additional term , \ $\beta_{0} (l_{pl}^2 (\Delta p)^2/ \hbar)
$ has its origin on the very nature of spacetime at the Planck
energy scale. It was shown in \cite{1} that the simplest GUP
relation which implies the appearance of a nonzero minimal
uncertainty $\Delta x_{0}$ in position has the form
\begin{equation}
\Delta x \Delta p\ \geq\ \frac{\hbar}{2}\ \Big(\ 1 + \beta (\Delta
p)^2 + \beta \langle p\rangle^2\ \Big)\,,
\end{equation}
where $\beta$ is the GUP parameter defined as $\beta = \beta_0
/(M_{pl}c)^2 = \beta_0 l_{pl}^2/\hbar^2 $, \ and $ \ M_{pl}c^2
\approx 10^{19}$ GeV is the 4-dimensional fundamental scale. It is
normally assumed that $\beta_0 $, which is a dimensionless number,
is not far from unity, that is, $\beta_0 \approx 1 $ (see for
instance \cite{23}-\cite{28}). At energies much below the Planck
energy, the extra term in right hand side of (2) would be
irrelevant, which means $\beta \rightarrow 0 $\ \ and the standard
HUP relation is recovered. Instead, approaching the Planck energy
scale, this term becomes relevant and, as has been said, it is
related to the minimal measurable length. From an string theory
viewpoint, since a string cannot probe distances smaller than its
length, existence of a minimal observable length is natural.

Since for any pair of observables \textbf{A} and \textbf{B} (which
are represented as symmetric operators on a domain of $\textbf{A}^2$
and $\textbf{B}^2$) one has
$$ \Delta \textbf{A} \Delta \textbf{B}\
 \geq\ \frac{\hbar}{2} \mid\langle\ [\textbf{A},\textbf{B}]\
\rangle\mid,$$ one finds the following algebraic structure
\begin{equation}
[x,p] = i\hbar\ (1 + \beta p^2)\,.
\end{equation}
Following Ref.\cite{1}, we define position and momentum operators
for the GUP case as
\begin{equation}
X = x
\end{equation}
\begin{equation}
P = p\ ( 1 + \beta p^2)
\end{equation}
where \emph{x} and \emph{p} ensure the Jacobi identities, namely
$[x_i,p_j] =i\hbar \delta_{ij}$\,, $[x_i,x_j] = 0$\, and
$[p_i,p_j] = 0$. Now it is easy to show that X and P satisfy the
generalized uncertainty principle. We interpret $p$ as the
momentum operator at low energies which has the standard
representation in position space, i.e. $ p_j = \frac {\hbar}
{i}\frac {\partial} {\partial{x}_j}\ $, and $P$ as the momentum
operator at high energies, where has the generalized
representation in position space, i.e. $ P_j = \frac {\hbar} {i}
\frac {\partial} {\partial{x}_j} \Big[ 1 + \beta(\frac {\hbar}
{i}\ \frac {\partial} {\partial{x}_j})^2\Big].$

\subsection{GUP with minimal length and Maximal momentum}

Magueijo and Smolin have shown that in the context of the doubly
special relativity (DSR), a test particles' momentum cannot be
arbitrarily imprecise and therefore there is an upper bound for
momentum fluctuations \cite{33}-\cite{35}. Then it has been shown
that this may lead to a maximal measurable momentum for a test
particle \cite{36}. In this framework, the GUP that predicts both
a minimal observable length and a maximal momentum can be written
as follows \cite{23}-\cite{28}
\begin{equation}
\Delta x \Delta p\ \geq\ \frac{\hbar}{2}\ \Big(\ 1 - 2\alpha \langle
p\rangle +\ 4\alpha^2 \langle p^2\rangle\ \Big)\,.
\end{equation}
In this framework the following algebraic structure can be deduced
(see \cite{28} for details)
\begin{equation}
[x,p] = i\hbar\Big(1 - \alpha p + 2\alpha^2 p^2\Big)
\end{equation}
where $\alpha $ is the GUP parameter in the presence of both minimal
length and maximal momentum and $\alpha = \alpha_0 /M_{pl}c =
\alpha_0 l_{pl}/\hbar $ . We note that the constants $\alpha$ and
$\beta$ are related through dimensional analysis with the expression
[$\alpha^{2}$] = [$\beta$]. Similar to the minimal length case, we
can define \cite{28}
\begin{equation}
\textbf{X} = x
\end{equation}
\begin{equation}
\textbf{P} = p\ \Big( 1 - \alpha p + 2\alpha^2 p^2\Big)
\end{equation}
where, as before $x$ and $p$ satisfy the canonical commutation
relations via the Jacobi identity, and, $\textbf{X}$ and
$\textbf{P}$ satisfy the generalized commutation relation in the
presence of minimal length and Maximal momentum
\begin{equation}
[\textbf{X},\textbf{P}] = i\hbar\Big( 1 - \alpha p + 2\alpha^2
p^2\Big).
\end{equation}
In comparison with the previous subsection, here there is an extra,
first order term in particle's momentum which has its origin on the
existence of a maximal momentum. This term is the source of
differences between our analysis of the Hilbert space representation
and the seminal work presented in Ref.\cite{1}. Before treating our
main problem, here we digress for a moment to show how maximal
momentum arises in this setup. The absolute minimal measurable
length in our setup is given by $ \Delta x_{min}(<p>=0)\equiv\Delta
x_0=2\alpha\hbar $ (see Eq.(22) below). Due to duality of position
and momentum operators, it is reasonable to assume $\Delta
x_{min}\propto\Delta p_{max}$. Now with
\begin{equation}
\Delta x\Delta p=
\frac{\hbar}{2}\Big(1-2\alpha<p>+4\alpha^2<p^2>\Big)
\end{equation}
in the boundary of the allowed region and setting $<p>=0$ to obtain
absolute maximal momentum, we arrive at
\begin{equation}
\Delta x \Delta p=\frac{\hbar}{2}\Big(1+4\alpha^2<p^2>\Big).
\end{equation}
Since $\Delta p=\sqrt{<p^2>-<p>^2} $, we find
\begin{equation}
\Delta x \Delta p=\frac{\hbar}{2}\Big(1+4\alpha^2(\Delta
p)^2+4\alpha^2<p>^2\Big).
\end{equation}
This results in
\begin{equation}
(\Delta p)^2-\frac{\Delta x}{2\alpha^2\hbar}\Delta
p+\frac{1}{4\alpha^2}=0.
\end{equation}
So we find
\begin{equation}
(\Delta p_{max})^2-\frac{\Delta x_{min}}{2\alpha^2\hbar}\Delta
p_{max}+\frac{1}{4\alpha^2}=0.
\end{equation}
Now using the value of $\Delta x_{min}$, we find
\begin{equation}
(\Delta p_{max})^2-\frac{1}{\alpha}\Delta
p_{max}+\frac{1}{4\alpha^2}=0.
\end{equation}
The solution of this equation is
\begin{equation}
\Delta p_{max} =\frac{1}{2\alpha}.
\end{equation}
As a nontrivial assumption we assume this is the maximal momentum in
our setup. We use this value in our forthcoming arguments.

In which follows, we reconsider the issue of Hilbert space
representation of quantum mechanics in the line of Ref.\cite{1}
but with new ingredient coming from existence of a maximal
momentum. We show that there are a lots of new implications in
this framework.

\section{Hilbert space representation}

As a nontrivial assumption, we assume the minimal observable
length to be also minimal, nonzero uncertainty in position.
Therefore, we have no longer a Hilbert space representation on
position space wave functions of the ordinary quantum mechanics.
This is because there is no more physical state which is a
position eigenstate $|x\rangle $, since an eigenstate would have
zero uncertainty in position. This means that we must construct a
new Hilbert space representation which is compatible with our
commutation relation in GUP (11). Fortunately, by neglecting the
presence of a minimal uncertainty in momentum, there still would
exist a continuous momentum space representation, which means that
we can explore the physical implications of the minimal length by
working with the convenient representation of the commutation
relations on momentum space wave functions.

\subsection{Some consequences in momentum space}

In this subsection we consider the momentum space representation. To
obtain a minimum measurable uncertainty in position, the inequality
(7) on the boundary of the allowed region gives
\begin{equation}
\Delta x \Delta p\ \ =\ \frac{\hbar}{2}\Big(\ 1 - 2\alpha \langle
p\rangle +\ 4\alpha^2 \langle p^2\rangle\Big)
\end{equation}
Using $\langle p^2\rangle = (\Delta p)^{2}+ \langle p\rangle^{2}$,
this relation can be rewritten as a second order equation for
$\Delta p$. The solutions for $\Delta p$ are as follows
\begin{equation}
\Delta p = \frac{\Delta
x}{{4\alpha}^{2}{\hbar}}\pm\sqrt{\bigg(\frac{\Delta
x}{{4\alpha}^{2}{\hbar}}\bigg)^2-\frac{\langle p\rangle}{2
\alpha}\Big(2\alpha\langle p\rangle - 1\Big) -
\frac{1}{4\alpha^2}}\,.
\end{equation}
The reality of solutions gives the following minimum value for
$\Delta x $
\begin{equation}
\Delta x_{{\min}} \ ({\it \langle p \rangle}) = 2 \alpha \hbar\
\sqrt {1 - 2 \alpha \langle p\rangle + 4 \alpha^2 \langle p\rangle^2
}\,.
\end{equation}
Therefore the absolutely smallest uncertainty in position, where
$\langle p\rangle = 0$, would be
\begin{equation}
\Delta x_0 = 2 \alpha \hbar .
\end{equation}
Now, in our momentum space, we take operators \textbf{P} and
\textbf{X} in the form
\begin{equation}
\textbf{P} = p
\end{equation}
\begin{equation}
\textbf{X} = \Big( 1 - \alpha p + 2\alpha^2 p^2\Big)\ x
\end{equation}
where \emph{x} = $i \hbar \frac{\partial}{\partial p} $ . Then by
operating on momentum space wave function, we have
\begin{equation}
\textbf{P}\varphi(p) = p\varphi(p)
\end{equation}
\begin{equation}
\textbf{X}\varphi(p) = i\hbar\Big( 1 - \alpha p + 2\alpha^2 p^2
\Big)\ \frac {\partial}{\partial p} \varphi(p)\,.
\end{equation}
The scalar product in this representation should be modified due
to the presence of the additional factor $(1 - \alpha p +
2\alpha^2 p^2 )\equiv G_{Mm}(p)$\, and existence of the maximal
momentum as
\begin{equation}
\langle \Phi | \varphi \rangle\ =\ \int_{-P_{pl}}^{+P_{pl}}\ \frac
{dp}{1 - \alpha p + 2\alpha^2 p^2}\ \Phi^*(p)\ \varphi(p)\,.
\end{equation}
We note that in KMM (Kempf {\it et al.}) formalism, since there is
no restriction on particle's momentum, the integrals are calculated
from $-\infty$ to $+\infty$. Here, the presence of the term \,
$-\alpha p$\, in $G_{Mm}(p)$ implies the existence of a maximal
particle's momentum (the Planck momentum, $ P_{pl} \equiv M_{pl} c
$)  which affects the scalar product as we see in Eq.(27). This fact
requires a reconsideration of the KMM formulation of the Hilbert
space representation. In this framework, the identity operator would
be represented as
\begin{equation}
\textbf{1}\ =\ \int_{-P_{pl}}^{+P_{pl}}\  \frac { dp}{1 - \alpha p +
2\alpha^2 p^2}\ |p\rangle \langle p|\,,
\end{equation}
and the scalar product of the momentum eigenstates changes to
\begin{equation}
\langle p | p' \rangle\ =\ \Big( 1 - \alpha p + 2\alpha^2 p^2 \Big)\
\delta (p-p')\,.
\end{equation}
The existence of maximal particle's momentum in addition to
minimal observable length, has several new implications on the
Hilbert space representation that we are going to study some of
them in forthcoming sections.

\subsection{Formal eigenstates of position operator in momentum space}

The position operator acting on momentum space eigenstates gives
\begin{equation}
\textbf{X} \varphi_\zeta(p) = \zeta\ \varphi_\zeta(p)\,,
\end{equation}
where by definition $ \varphi_{\zeta}(p) = \langle p | \zeta
\rangle $ is a formal position eigenstate and $|\zeta\rangle $ is
an arbitrary state. So, we find
\begin{equation}
i \hbar\Big( 1 - \alpha p + 2\alpha^2 p^2 \Big)\ \frac {\partial
\varphi_\zeta(p)}{\partial p} = \zeta\ \varphi_\zeta(p)\,.
\end{equation}
By solving this differential equation, we obtain the formal
position eigenvectors in the following form
\begin{eqnarray}
\varphi_\zeta(p)=\varphi_\zeta(0)\exp\Big[-i\frac{2\zeta}{\alpha
\hbar\sqrt{7}}\Big\{\tan^{-1}(\frac{1}{\sqrt
{7}})\ +\ \ \ \ \ \ \ \ \ \ \ \nonumber\\
\tan^{-1} \Big(\frac{4\alpha p - 1}{\sqrt
{7}}\Big)\Big\}\Big]\,.\hspace{0.5cm}
\end{eqnarray}
Then by normalization, $ \langle \varphi|\varphi\rangle =
\textbf{1} $ we have
$$\textbf{1}\ =\ \int_{-P_{pl}}^{+P_{pl}}\ \frac
{1}{1 - \alpha p + 2\alpha^2 p^2}\ {\varphi^{*}_\zeta}(p)\
\varphi_\zeta(p)\ dp\ $$
\begin{equation}
=\varphi_\zeta(0)\ {\varphi^{*}_\zeta}(0)
\int_{-P_{pl}}^{+P_{pl}}\ \frac {dp}{1 - \alpha p + 2\alpha^2
p^2}\,,
\end{equation}
so, we find
\begin{eqnarray}
\varphi_\zeta(0)=\sqrt{\frac{\alpha\sqrt{7}}{2}}\bigg[\tan^{-1}\Big(
\frac {4 \alpha P_{pl} - 1} {\sqrt
{7}}\Big)+\ \ \ \ \ \ \ \ \ \ \ \ \ \ \ \ \ \ \nonumber\\
\tan^{-1}\Big(\frac {4 \alpha P_{pl} + 1} {\sqrt
{7}}\Big)\bigg]^{-1/2}.\hspace{5mm}
\end{eqnarray}
Thus the formal position eigenvectors in momentum space would be
$$\varphi_\zeta(p) =
\sqrt{\frac{\alpha\sqrt{7}}{2}}\bigg[\tan^{-1}\Big(\frac{4\alpha
P_{pl}-1}{\sqrt{7}}\Big)+\ \ \ \ \ \ \ \ \ \ \ \ \ \ \ \ \ \ \ $$
\begin{equation}
\tan^{-1}\Big(\frac{4\alpha P_{pl} + 1}
{\sqrt{7}}\Big)\bigg]^{-\frac{1}{2}}e^{-i\frac{2\zeta}{\alpha
\hbar \sqrt{7}}\Big\{\tan^{-1}(\frac{1}{\sqrt{7}})+
\tan^{-1}\Big(\frac {4\alpha p-1}{\sqrt{7}}\Big)\Big\}}\,.
\end{equation}
This is the generalized, momentum space eigenstate of the position
operator in the presence of both a minimal length and a maximal
momentum. Now we calculate the scalar product of the formal
position eigenstates as

$$ \langle \varphi_{\zeta'} | \varphi_\zeta \rangle\ =
\int_{-P_{pl}}^{+P_{pl}}\ \frac {1}{1 - \alpha p + 2\alpha^2 p^2}\
{\varphi^{*}_{\zeta'}}(p)\ \varphi_\zeta(p)\ dp  $$

$$=\frac{\alpha\sqrt{7}}{2}\bigg[\tan^{-1}\Big(\frac{4\alpha P_{pl}-1}{\sqrt{7}}\Big)+\tan^{-1}\Big(\frac{4\alpha P_{pl}+
1}{\sqrt{7}}\Big)\bigg]^{-1}\times$$

$$\hspace{1.5cm}\int_{-P_{pl}}^{+P_{pl}}\frac{e^{-i
\frac{2(\zeta-\zeta')}{\alpha\hbar\sqrt{7}}\Big(\tan^{-1}(\frac{1}
{\sqrt{7}})+\tan^{-1}(\frac{4\alpha p-1}{\sqrt{7}})\Big)}}{1 -
\alpha p + 2\alpha^2 p^2}dp $$

$$=\frac{\alpha\sqrt{7}}{2}\Big[\tan^{-1}(\frac{4\alpha P_{pl}-1}{\sqrt{7}})+\tan^{-1}(\frac{4
\alpha P_{pl} + 1}{\sqrt{7}})\Big]^{-1}\times$$

\begin{equation}
e^{-i\frac{2(\zeta-\zeta')}{\alpha\hbar\sqrt{7}}\tan^{-1}(\frac{1}
{\sqrt{7}})}\int_{-P_{pl}}^{+P_{pl}}\frac{e^{-i\frac{2(\zeta-\zeta')}{\alpha\hbar\sqrt{7}}
\tan^{-1}(\frac{4\alpha p - 1}{\sqrt{7}})}}{1 - \alpha p +
2\alpha^2 p^2}dp\,,
\end{equation}\\

where by solving this integral, we find the following result
\begin{eqnarray}
\langle\varphi_{\zeta'}|\varphi_\zeta\rangle=i\Upsilon_0\bigg[e^{-i
\frac {2(\zeta - \zeta')} {\alpha \hbar \sqrt {7}}
\bigg(\tan^{-1}( \frac {4 \alpha P_{pl} - 1} {\sqrt {7}})\ +\
\tan^{-1}( \frac {1} {\sqrt {7}})\bigg)}-\nonumber\\
e^{i \frac {2(\zeta - \zeta')} {\alpha \hbar \sqrt {7}}
\bigg(\tan^{-1}( \frac {4 \alpha P_{pl} + 1} {\sqrt {7}})\ -\
\tan^{-1}( \frac {1} {\sqrt {7}})\bigg)}\bigg]\,,\hspace{7mm}
\end{eqnarray}
where by definition
\begin{eqnarray}
\Upsilon_0\equiv\frac{\sqrt{7}\alpha\hbar}{2(\zeta-\zeta')}\bigg[\tan^{-1}\Big(\frac{4\alpha
P_{pl}-1}{\sqrt {7}}\Big)+\ \ \ \ \ \ \ \ \ \ \ \ \ \ \ \ \nonumber\\
\tan^{-1}\Big(\frac{4\alpha P_{pl}+1}
{\sqrt{7}}\Big)\bigg]^{-1}.\hspace{4mm}
\end{eqnarray}\\
We note that since $i = e^{i \pi/2}$, Eq.(37) can be rewritten as
\begin{widetext}
\begin{equation}
\langle \varphi_{\zeta'} | \varphi_\zeta \rangle\ = \Upsilon_0\
e^{-i \frac {2(\zeta - \zeta')}{\alpha \hbar \sqrt {7}}\\tan^{-1}(
\frac {1} {\sqrt {7}})}\bigg[e^{-i\bigg(\frac {2(\zeta - \zeta')}
{\alpha \hbar \sqrt {7}} \tan^{-1}( \frac {4 \alpha P_{pl} - 1}
{\sqrt {7}})\ -\ \frac {\pi}{2}\bigg)} -\ e^{i\bigg(\frac {2(\zeta
- \zeta')} {\alpha \hbar \sqrt {7}} \tan^{-1}( \frac {4 \alpha
P_{pl} + 1} {\sqrt {7}})\ +\ \frac {\pi}{2}\bigg)}\bigg]\,.
\end{equation}
\end{widetext}
Figure 1 compares the behavior of $\langle \varphi_{\zeta'} |
\varphi_\zeta \rangle\ $ versus $\zeta - \zeta'$ in our and the
KMM frameworks. The scalar product $\langle \varphi_{\zeta'} |
\varphi_\zeta \rangle\ $ in our case has more broadening relative
to the KMM case.

\begin{figure}
\flushleft\leftskip-0em{\includegraphics[width=2.5in]{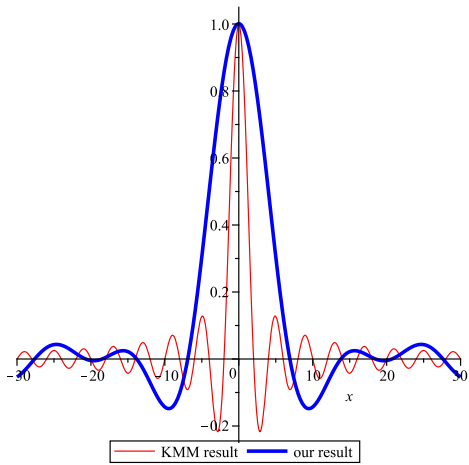}}\hspace{3cm}
\caption{\label{fig:1}Variation of $\langle \varphi_{\zeta'} |
\varphi_\zeta \rangle\ $ versus $\zeta - \zeta'$.}
\end{figure}

Now we calculate the expectation value of energy for these formal
position eigenvectors
\begin{eqnarray}
\langle\varphi_\zeta|\frac{p^2}{2m}|\varphi_\zeta\rangle=\hspace{35mm}\nonumber\\
\int_{-P_{pl}}^{+P_{pl}} \frac{1}{1-\alpha p + 2\alpha^2
p^2}\ {\varphi^{*}_\zeta}(p)\ \frac{p^2}{2m}\ \varphi_\zeta(p)\ dp\ =\hspace{10mm}\nonumber\\
\frac{\alpha\sqrt{7}}{4m}\bigg[\tan^{-1}\Big(\frac{4\alpha P_{pl}
- 1}{\sqrt{7}}\Big)+\tan^{-1}\Big(\frac{4\alpha P_{pl}+ 1}
{\sqrt{7}}\Big)\bigg]^{-1}\times\hspace{4mm}\nonumber\\
\int_{-P_{pl}}^{+P_{pl}}\frac{p^2}{1 - \alpha p + 2\alpha^2
p^2}dp\,.\hspace{10mm}
\end{eqnarray}

By solving this integral, we find
\begin{eqnarray}
\langle\frac{p^2}{2m}\rangle=\frac{\alpha\sqrt{7}}{224m\alpha^2}
{\Gamma}^{-1}\bigg[56\alpha P_{pl}+7\ln\Big(\frac{1-\alpha
P_{pl}+2\alpha^2 P_{pl}^2}{1+\alpha P_{pl}+2\alpha^2
P_{pl}^2}\Big)\nonumber\\
-\ 6\sqrt{7}\Gamma\bigg]\hspace{10mm}
\end{eqnarray}
where by definition
$$ \Gamma\ \equiv\ \tan^{-1}\Big( \frac {4 \alpha P_{pl} - 1} {\sqrt {7}}\Big)\
+\ \tan^{-1}\Big(\frac {4 \alpha P_{pl} + 1} {\sqrt {7}}\Big)\,.
$$
As one can see by comparison with KMM formalism, unlike the KMM
case, the expectation value of energy in our case is no longer
divergent. Kempf {\it et al.} in Ref.\cite{1} with just minimal
length GUP, have stated that " the formal position eigenvectors $
|\varphi_\zeta \rangle $ are not physical states, because they are
not in the domain of \textbf{P} which physically means that they
have infinite uncertainty in momentum and in particular also
infinite energy ". As we have shown, in the presence of both
minimal length and maximal momentum there is no divergency in
energy spectrum. Kempf {\it et al.} have indicated also that "
vectors $|\psi\rangle $ that have a well-defined uncertainty in
position $\Delta x_{|\psi\rangle}$ which is inside the forbidden
gap $0\leq \Delta x_{|\psi\rangle}\leq\Delta x_0$ cannot have
finite energy". As we have shown, there is no longer divergency in
energy for the formal position eigenvectors and
$|\varphi_\zeta\rangle $ have no longer infinite uncertainty in
momentum. Nevertheless, we note that $|\varphi_\zeta\rangle $ are
not physical states in our case too since $\Delta x_0\neq 0$.

This is an important implication of the presence of a maximal
momentum which restricts the maximum value of the particle's
momentum to an upper bound of  $ P_{pl} $.\\

\section{ Maximal localization and its consequences}

Due to the presence of the minimal length, $l_{pl}=\Delta
x_0=2\alpha\hbar$, one cannot probe distances less than the Planck
length. So, the very notion of the spacetime manifold should be
reconsidered due to the finite resolution of the spacetime points.
If we treat the problem in a realistic manner, we are forced to
introduce the maximal localization states that are localized just
up to the Planck length and there is no further localization
possible in essence. In what follows, we concern on physical
states that are maximally localized around a classical spacetime
point. There is no longer the very notion of localization like
that was in ordinary quantum mechanics.

\subsection{ Some analysis on maximal localization states}

Now we consider the states $|\varphi^{ml}_\varepsilon \rangle $ of
\emph{maximal localization} around a position $\varepsilon$\, that
$\varepsilon \geq l_p$
\begin{equation}
\langle \varphi^{ml}_\varepsilon | \textbf{X} |
\varphi^{ml}_\varepsilon \rangle = \varepsilon\,.
\end{equation}
As Kempf \emph{et al.} mentioned in Ref.\cite{1}, from the
positivity of the norm, that is,
\begin{equation}
\| \bigg( \textbf{X} - \langle \textbf{X} \rangle + \frac {\langle
[\textbf{X},\textbf{P}] \rangle} {2(\Delta \textbf{P})^2}(\textbf{P}
- \langle \textbf{P} \rangle) \bigg)\ |\varphi \rangle \|\ \geq\
0\,,
\end{equation}
on the boundary of the physically allowed region, we have
\begin{equation}
\bigg( \textbf{X} - \langle \textbf{X} \rangle + \frac {\langle
[\textbf{X},\textbf{P}] \rangle} {2(\Delta \textbf{P})^2}(\textbf{P}
- \langle \textbf{P} \rangle) \bigg)\ |\varphi \rangle\ =\ 0\,.
\end{equation}
Using Eqs.(25) and (26), the differential equation in momentum space
corresponding to (44) is in the following form
\begin{eqnarray}
\Bigg(i\hbar\Big(1-\alpha p+2\alpha^2
p^2\Big)\frac{\partial}{\partial p}-\langle\textbf{X}\rangle+\hspace{25mm}\nonumber\\
i\hbar\frac{1+2\alpha^2(\Delta p)^2+2\alpha^2\langle
p\rangle^2-\alpha\langle p\rangle}{2(\Delta p)^2}\Big(p-\langle
p\rangle\Big)\Bigg)|\varphi\rangle=0\,.\hspace{10mm}
\end{eqnarray}
By solving this differential equation and taking into account that
$\langle\textbf{X}\rangle=\varepsilon $\,, we obtain the states of
maximal localization as follows
\begin{widetext}
\begin{eqnarray}
\varphi^{ml}_\varepsilon(p)=\Phi
e^{\frac{-2}{\alpha\hbar\sqrt{7}}\bigg[\frac{\hbar}{2(\Delta
p)^2}\Big(\frac{1}{4 \alpha}-\langle
p\rangle\Big)\Big(1+2\alpha^2(\Delta p)^2+2\alpha^2\langle
p\rangle^2-\alpha\langle p\rangle\Big)+
i\varepsilon\bigg]\Big(\tan^{-1}(\frac{1}{\sqrt{7}})+\tan^{-1}(\frac
{4\alpha p-1}{\sqrt{7}})\Big)}
\end{eqnarray}
\end{widetext}

where

$$ \Phi\ \equiv\ \varphi^{ml}_\varepsilon(0)\ (1 -
\alpha p + 2\alpha^2 p^2)^{- \frac{1 + 2 \alpha^2 (\Delta p)^2 + 2
\alpha^2 \langle p \rangle^2 - \alpha \langle p \rangle}{8
\alpha^2 (\Delta p)^2}}\,.$$

For $\langle p\rangle=0$ and $\Delta p=\frac{1}{2\alpha}$ that are
corresponding to the states of absolutely maximal localization and
critical momentum uncertainty, the minimal position uncertainty is
recovered. The corresponding states are given as

\begin{eqnarray}
\varphi^{ml}_\varepsilon(p)=\frac{\varphi^{ml}_\varepsilon(0)\
e^{-\frac{3}{2\sqrt{7}}\Big(\tan^{-1}(\frac{1}{\sqrt{7}})
+\tan^{-1}(\frac{4\alpha p-1}{\sqrt{7}})\Big)}}{(1-\alpha
p+2\alpha^2 p^2)^{\frac{3}{4}}}\times\hspace{8mm}\nonumber\\
e^{-i\frac{2\varepsilon}{\alpha\hbar\sqrt{7}} \Big(\tan^{-1}(\frac
{1}{\sqrt{7}})+\tan^{-1}(\frac{4\alpha
p-1}{\sqrt{7}})\Big)}\,.\hspace{10mm}
\end{eqnarray}
By normalization to unity, $\langle
\varphi^{ml}_\varepsilon|\varphi^{ml}_\varepsilon\rangle
=\textbf{1}$, we find

$$\textbf{1}=\int_{-P_{pl}}^{+P_{pl}}\frac{1}{1-\alpha p+2\alpha^2
p^2}{\varphi^{ml}_\varepsilon}^*(p)\varphi^{ml}_\varepsilon(p)\
dp\ =\hspace{5mm} $$
\begin{equation}
\varphi^{ml}_\varepsilon(0){\varphi^{ml}_\varepsilon}^*(0)\int_{-P_{pl}}^{+P_{pl}}\frac
{e^{-\frac{3}{\sqrt{7}}\Big(\tan^{-1}(\frac{1}{\sqrt{7}})+\tan^{-1}(\frac
{4\alpha p-1}{\sqrt{7}})\Big)}}{(1-\alpha p+2\alpha^2
p^2)^{\frac{5}{2}}} dp\,,
\end{equation}\\
which gives
\begin{eqnarray}
\varphi^{ml}_\varepsilon(0)=\sqrt{6\alpha}\Big[\sqrt{8}
e^{\eta\tan^{-1}(\eta)}-\hspace{35mm}\nonumber\\
e^{-\eta\tan^{-1}(\eta/3)}\Big]^{-1/2}
e^{\frac{\eta}{2}\tan^{-1}(\eta/3)}\hspace{5mm}
\end{eqnarray}
where, $ \eta\equiv\frac{4\alpha P_{pl}-1}{\sqrt{7}}=\frac
{3}{\sqrt{7}}$, since, $P_{pl}=\frac{1}{2\alpha}$. Therefore, the
momentum space wave functions $\varphi^{ml}_\varepsilon(p)$ of the
states that are maximally localized around a position
$\varepsilon$, $ \langle \textbf{X} \rangle = \varepsilon $, are
in the following form
\begin{eqnarray}
\varphi^{ml}_\varepsilon(p)=\frac{\sqrt{6\alpha}\Big[\sqrt{8}
e^{\eta\tan^{-1}(\eta)}-e^{-\eta\tan^{-1}(\frac{\eta}{3})}\Big]^{-\frac{1}{2}}}
{(1-\alpha p+2\alpha^2 p^2)^{\frac{3}{4}}}\times\hspace{10mm}\nonumber\\
e^{-\frac{\eta}{2} \tan^{-1}(\frac{4\alpha p-1}{\sqrt{7}})}
e^{-i\frac{2\varepsilon}{\alpha\hbar\sqrt{7}}\Big(\tan^{-1}(\frac{\eta}{3})
+\tan^{-1}(\frac {4\alpha p-1}{\sqrt{7}})\Big)}.\hspace{8mm}
\end{eqnarray}
One can see easily the difference between this result and the
corresponding wave function obtained by Kempf {\it et al.} \cite{1}.
This difference has its origin on the presence of the term that is
first order of momentum in the right hand side of Eq.(24), which
implies the existence of a maximal momentum.

Now we calculate the expectation value of energy for this maximal
localization state wave function

$$\langle\varphi^{ml}_\varepsilon|\frac{p^2}{2m}|\varphi^{ml}
_\varepsilon\rangle\ =\hspace{60mm}$$
$$\int_{-P_{pl}}^{+P_{pl}}\frac{1}{1-\alpha p+2\alpha^2 p^2}\
{\varphi^{ml}_\varepsilon}^*(p) \frac{p^2}{2m} \varphi^{ml}
_\varepsilon(p)\ dp\ =\hspace{3cm}$$
\begin{eqnarray}
\frac{3\alpha}{m}\bigg[\sqrt{8}e^{\eta\tan^{-1}(\eta)}-
e^{-\eta\tan^{-1}(\eta/3)}\bigg]^{-1}\times\hspace{2cm}\nonumber\\
\int_{-P_{pl}}^{+P_{pl}}\frac{e^{-\bigg(\eta\tan^{-1}(\frac{4\alpha
p-1}{\sqrt{7}})\bigg)}}{(1-\alpha p+2\alpha^2
p^2)^{5/2}}p^2dp.\hspace{0.7cm}
\end{eqnarray}\\
Solving this integral, we obtain
\begin{equation}
\langle\frac{p^2}{2m}\rangle=\frac{1}{4m\alpha^2}\bigg(\frac{1}{2}\frac{\sqrt
{2}e^{\eta\tan^{-1}(\eta)}-3e^{-\eta\tan^{-1}(\eta/3)}}{\sqrt{8}e^{\eta\tan^{-1}(\eta)}
-e^{-\eta\tan^{-1}(\eta/3)}}\bigg)\,.
\end{equation}\\
This result differs with Kempf {\it et al.} result by the factor
in the parenthesis. With $ \eta = \frac {3}{\sqrt {7}}$, since
$$ \sqrt {2}\ e^{\ \eta \tan^{-1}( \eta)} -\ 3\ e^{\ -\eta
\tan^{-1}( \eta/3)}\ \approx\ 1.71\,, $$ and
$$ \sqrt {8}\ e^{\ \eta \tan^{-1}( \eta)} -\ e^{\
-\eta \tan^{-1}( \eta/3)}\ \ \ \approx\ 6.75\,, $$ the expectation
value of energy for our maximal localization state wave function
would be
\begin{equation}
\langle \frac {p^2}{2 m} \rangle\ \approx\ \frac {1} {32m
\alpha^2}\,.
\end{equation}
Comparing this result with $\frac{1}{2m \beta} $ obtained by KMM
(with $\beta\approx 2\alpha^{2}$), shows the important role played
by the maximal momentum in this setup. The scalar product of the
maximal localization states is as follows
$$\langle\varphi^{ml}_{\varepsilon'}|\varphi^{ml}_\varepsilon\rangle
=$$
\begin{eqnarray}
C_0\int_{-P_{pl}}^{+P_{pl}}\frac{e^{-\bigg(\eta\tan^{-1}(\frac
{4\alpha p-1}{\sqrt{7}})\bigg)}}{(1-\alpha p+2\alpha^2
p^2)^{5/2}}\times\hspace{6.5cm}\nonumber\\
e^{-i\frac{2(\varepsilon-\varepsilon')}{\alpha\hbar\sqrt
{7}}\bigg(\tan^{-1}(\frac{\eta}{3})+\tan^{-1}(\frac{4\alpha p-1}
{\sqrt{7}})\bigg)}dp=\hspace{3cm}\nonumber\\
C_0e^{-i\frac{2(\varepsilon-\varepsilon'
)}{\alpha\hbar\sqrt{7}}\tan^{-1}(\frac{\eta}{3})}\int_{-P_{pl}}^{+P_{pl}}\frac
{e^{-\bigg(\eta \tan^{-1}(\frac{4\alpha p-1}{\sqrt{7}})\bigg)}}{(1
-\alpha p+2\alpha^2 p^2)^{5/2}}\times\hspace{4cm}\nonumber\\
e^{-i\frac {2(\varepsilon-\varepsilon')}{\alpha\hbar\sqrt
{7}}\bigg(\tan^{-1}(\frac{4\alpha p-1}{\sqrt{7}})\bigg)}
dp=\hspace{3.3cm}
\end{eqnarray}
\begin{equation}
\frac{32\sqrt{2}}{49\alpha}C_0e^{-i\frac{2(\varepsilon-\varepsilon')}{\alpha
\hbar\sqrt{7}}\tan^{-1}(\frac{\eta}{3})}\int_{-\tan^{-1}(3/\sqrt
{7})}^{+\tan^{-1}(1/\sqrt{7})}\frac{e^{\frac{-3u}{\sqrt{7}}}e^{-i\frac{2(\varepsilon
-\varepsilon')u}{\alpha\hbar\sqrt{7}}}}{\bigg(1+\tan^2(u)\bigg)^{3/2}}du
\end{equation}

\begin{widetext}
\begin{equation}
=\Lambda C_0\bigg\{\bigg[9\alpha\hbar\Delta^2-
96{\alpha}^{3}{\hbar}^{3}+i({\Delta}^{3}
-46{\alpha}^{2}{\hbar}^{2}\Delta)\bigg]e^\Sigma+
2\sqrt{2}\bigg[12{\alpha}^{3}{\hbar}^{3}-3\alpha\hbar\Delta^2+i(
16{\alpha}^{2}{\hbar}^{2}\Delta-{\Delta}^{3})\bigg]
e^{-i\Omega}\bigg\}\hspace{0.9cm}
\end{equation}
\end{widetext}
\begin{widetext}
$$$$
\hspace{7.3cm}where by definition
$$$$
$$\Lambda\equiv-\frac{\hbar}{2\sqrt{2}}\frac{e^{-\frac{\eta}{2}
(\pi-2\tan^{-1}(\sqrt{7}))}e^{-i\frac{2\Delta}{\alpha\hbar\sqrt
{7}}\tan^{-1}(\frac{\eta}{3})}}{72\alpha^4\hbar^4-31\alpha^2\hbar^2\Delta^2
+\Delta^4+i(66\alpha^3\hbar^3\Delta-6\alpha\hbar\Delta^3)}$$
$$ C_0\ \equiv\ 6 \alpha\ \bigg[ \sqrt {8}\ e^{\ \eta \tan^{-1}( \eta)} -\ e^{\
-\eta \tan^{-1}(\frac{\eta}{3})} \bigg]^{-1} $$
$$ \Sigma\ \equiv\ \frac{3 \alpha\hbar\tan^{-1}(\sqrt {7})\ + i\ 2 \Delta \tan^{-1}(\eta)
}{ \alpha\hbar\sqrt {7}}$$
$$ \Omega\ \equiv\ \frac{\Delta}{\alpha\hbar\sqrt {7}}\ \bigg( \pi - 2\tan^{-1}(\sqrt {7}) \bigg) $$
$$ u\ \equiv\ \tan^{-1}( \frac {4 \alpha p - 1} {\sqrt {7}}) $$
$$ \Delta\ \equiv\ \varepsilon - \varepsilon'$$
$$ \eta \equiv\ \frac{3}{\sqrt {7}}\,.$$
\end{widetext}

As in the KMM case, there is no mutual orthogonality of the
maximal localization states in this case too. This is a
consequence of the spacetime fuzziness at the Planck scale. Figure
$2$ shows the behavior of $\langle \varphi^{ml}_{\varepsilon'} |
\varphi^{ml}_\varepsilon \rangle\ $ versus $\varepsilon -
\varepsilon'$.

\begin{figure}
\flushleft\leftskip-0em{\includegraphics[width=2.5in]{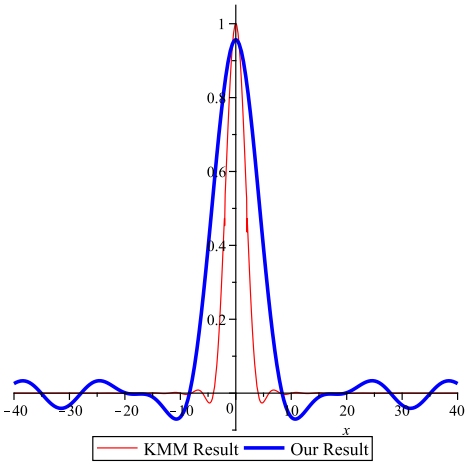}}\hspace{3cm}
\caption{\label{fig:2}Variation of $\langle
\varphi^{ml}_{\varepsilon'} | \varphi^{ml}_\varepsilon \rangle\ $
versus $\varepsilon - \varepsilon'$.}
\end{figure}

\subsection{ Quasiposition representation}

Here we consider the concept of quasi position wave function by
projecting arbitrary states on maximally localized states to obtain
the probability amplitude for the particle being maximally localized
around a position. We take $ | \phi \rangle\ $ as an arbitrary
state, then the probability amplitude on maximal localization states
around the position $ \varepsilon $ is $ \langle
\varphi^{Ml}_\varepsilon |\phi \rangle\ $ that we introduce it as
the state's quasiposition wavefunction $ \phi(\varepsilon) $. The
transformation of a state's wavefunction in the momentum
representation into its quasiposition wave function now would be
\begin{eqnarray}
\phi(\varepsilon)=\sqrt{C_0}\int_{-P_{pl}}^{+P_{pl}}\frac
{e^{-\bigg(\frac{\eta}{2}\tan^{-1}(\frac{4\alpha
p-1}{\sqrt{7}})\bigg)}}{(1-\alpha p+2\alpha^2
p^2)^{7/4}}\times\hspace{3cm}\nonumber\\
e^{i\frac{2\varepsilon}{\alpha\hbar\sqrt
{7}}\Big(\tan^{-1}(\frac{\eta}{3})+\tan^{-1}(\frac{4\alpha p-1}
{\sqrt{7}})\Big)}\phi(p)dp\,.\hspace{0.7cm}
\end{eqnarray}
This transformation that maps momentum space wave functions into
quasiposition space wave functions is the generalization of the
Fourier transformation.

Similar to the ordinary quantum mechanics case, we can write

\begin{equation}
e^{i \frac{2 \varepsilon}{\alpha \hbar \sqrt {7}}
\Big(\tan^{-1}(\frac{\eta}{3})\ +\ \tan^{-1}(\frac {4 \alpha p - 1}
{\sqrt {7}})\Big)}\ \equiv\ e^{i K \varepsilon}
\end{equation}
so that

\begin{equation}
K\ \equiv\ \frac{2}{\alpha \hbar \sqrt {7}}
\bigg(\tan^{-1}(\frac{\eta}{3})\ +\ \tan^{-1}(\frac {4 \alpha p - 1}
{\sqrt {7}})\bigg)
\end{equation}
is the modified wavenumber in our proposed setup. Therefore,
\begin{equation}
\lambda(p)\ =\ \frac {\pi \alpha \hbar \sqrt {7}} {
\tan^{-1}(\frac{\eta}{3})\ +\ \tan^{-1}(\frac {4 \alpha p - 1}
{\sqrt {7}}) }
\end{equation}
would be the modified wavelength for the quasiposition wavefunction
of physical states. Since $ \alpha \neq 0 $ and $p$ is finite
(limited to the Planck momentum), there is no wavelength smaller
than

\begin{equation}
\lambda_0\ =\ \lambda(P_{pl})\ =\ \frac {\pi \alpha \hbar \sqrt {7}}
{ \tan^{-1}(\frac{\eta}{3})\ +\ \tan^{-1}(\frac {4 \alpha P_{pl} -
1} {\sqrt {7}})}\,.
\end{equation}
Using the relation between energy and momentum $ E = p^2/2 m$ we
obtain
\begin{equation}
E(\lambda)\ =\ \frac {2}{m \alpha^2}\ \Bigg(\frac{\tan( \frac {\pi
\alpha \hbar \sqrt {7}} {\lambda})}{\sqrt{7} + \tan( \frac {\pi
\alpha \hbar \sqrt {7}} {\lambda})}\Bigg)^2\,.
\end{equation}
So, we find
\begin{equation}
E(\lambda_0)\ =\ \frac {{P_{pl}}^2}{2m}
\end{equation}
that is in agreement with ordinary quantum mechanics and doesn't
diverge. The importance of this result is that unlike the KMM
results (where the quasiposition wavefunctions in contrast to
ordinary quantum mechanics case had no longer arbitrarily fine
\textit{ripples}, because the energy of the short wavelength modes
were divergent), here similar to ordinary quantum mechanics case
those wavefunctions can have arbitrarily fine \textit{ripples}
because there is no longer divergency in energy for $\lambda_0 $.
This is an important outcome of our formalism with a GUP that
contains both minimal length and maximal momentum. If we set $ m
\approx M_{pl} $, then the energy of the short wavelength modes
will be the Planck energy \ $ E(\lambda_0) \approx E_{pl} $.

Note that Eq.(62) in the limit of $\alpha\rightarrow 0$ gives the
result
$$\lim_{\alpha\rightarrow 0}E_{\lambda}=\frac{p^{2}}{2m}$$ which is
reliable in the context of the correspondence principle.

\subsection{ Some consequences of the quasiposition representation}

By inverse Fourier transform of equation (57), we obtain
\begin{eqnarray}
\phi(p)=\Lambda_0\int_{-\infty}^{+\infty}\bigg(1-\alpha p +
2\alpha^2 p^2\bigg)^{3/4}e^{\bigg(\frac{\eta}{2}\tan^{-1}(\frac
{4\alpha p - 1}{\sqrt{7}})\bigg)}\times\nonumber\\
e^{-i\frac{2\varepsilon}{\alpha\hbar\sqrt{7}}\Big(\tan^{-1}(\frac{\eta}{3})
+\tan^{-1}(\frac{4\alpha p-1}{\sqrt{7}})\Big)}\phi(\varepsilon)
d\varepsilon\hspace{1cm}
\end{eqnarray}
which is transformation of a quasiposition wavefunction into a
momentum space wave function, where $\Lambda_0$ is given by
$$ \Lambda_0\ \equiv\  \bigg[ \frac{ \ \sqrt {8}\ e^{\ \eta \tan^{-1}( \eta)} -\ e^{\
-\eta \tan^{-1}( \eta/3)} }{24\ \pi^2\ \hbar^2 \alpha}\
\bigg]^{1/2}$$
with
$$ \eta = \frac {3}{\sqrt{7}}\,. $$
Note that the integral now is computed over $ -\infty $ to $ +\infty
$, because it is over $ d\varepsilon $ not $d\textit{p}$.

From Eq.(64) and adopting the same strategy as in ordinary quantum
mechanics, we can obtain the generalized form of the momentum
operator in quasiposition space. To do this end, we note that the
quasiposition representation is a generalized position space
representation respecting the fact that the notion of \emph{point}
is no longer the same as in classical physics or ordinary quantum
mechanics. Because now there is a minimum measurable length of the
order of the Planck length that restricts possible resolution of
spacetime points. In analogy with ordinary quantum mechanics, here
we consider a generalized plane wave profile in our generalized
space. Then, from Eq.(58) we can deduce
\begin{eqnarray}
\frac{\partial}{\partial\varepsilon}e^{i\frac{2\varepsilon}{\alpha\hbar\sqrt
{7}}\Big(\tan^{-1}(\frac{\eta}{3})+\tan^{-1}(\frac{4\alpha
p-1}{\sqrt{7}})\Big)} =\hspace{3cm}\nonumber\\
i\frac{2}{\alpha\hbar\sqrt{7}}\Big(\tan^{-1}(\frac{\eta}{3})+\tan^{-1}(\frac
{4\alpha p-1}{\sqrt{7}})\Big)\times\hspace{1cm}\nonumber\\
e^{i\frac{2\varepsilon}{\alpha\hbar\sqrt{7}}\Big(\tan^{-1}(\frac{\eta}{3})+\tan^{-1}(\frac
{4\alpha p-1}{\sqrt{7}})\Big)}\,.\hspace{0.8cm}
\end{eqnarray}
Formally this is equivalent to set
\begin{equation}
\frac{\alpha\sqrt{7}}{2}\frac{\hbar}{i}\frac{\partial}{\partial
\varepsilon} \equiv \Big(\tan^{-1}(\frac{\eta}{3})\ +\
\tan^{-1}(\frac {4 \alpha p - 1} {\sqrt {7}})\Big).
\end{equation}

Since
$$\tan \Big(\tan^{-1}(\frac{\eta}{3})\ +\ \tan^{-1}(\frac {4
\alpha p - 1} {\sqrt {7}})\Big) = \frac{\sqrt{7}\alpha p}{2-\alpha
p},$$

a simple calculation gives
\begin{equation}
\textbf{P}\ \equiv\ \frac{2}{\alpha}\ \frac{\tan\Big(\frac{\alpha
\sqrt {7}}{2}\ \frac{\hbar}{i}
\partial_\varepsilon\Big)}{\sqrt{7} + \tan\Big(\frac{\alpha \sqrt
{7}}{2}\ \frac{\hbar}{i} \partial_\varepsilon\Big)}\,.
\end{equation}
Using Eq.(26) and by the action of \textbf{X} on momentum space wave
functions (64), we derive the form of position operator in
quasiposition space as
\begin{equation}
\textbf{X}\ \equiv\ \varepsilon\ +\ i\ 6\alpha\hbar\
\frac{\tan\Big(\frac{\alpha \sqrt {7}}{2}\ \frac{\hbar}{i}
\partial_\varepsilon\Big)}{\sqrt{7} + \tan\Big(\frac{\alpha \sqrt
{7}}{2}\ \frac{\hbar}{i} \partial_\varepsilon\Big)}\,.
\end{equation}
So, the position and momentum operators $\textbf{X}$ and
$\textbf{P}$ in quasiposition space can be expressed in terms of the
multiplication and differentiation operators $ \varepsilon $ and $
\frac{\hbar}{i}\partial_\varepsilon\equiv p_{\varepsilon}$ which
obey the commutation relations of the ordinary quantum mechanics.
Note that in the limit of $\alpha\rightarrow 0$, we recover the
corresponding ordinary quantum mechanics results as
$$\lim_{\alpha\rightarrow
0}\textbf{P}=\frac{\hbar}{i}\frac{\partial}{\partial
\varepsilon}$$
and
$$\lim_{\alpha\rightarrow
0}\textbf{X}=\varepsilon$$
respectively.

We stress that as a result of quasiposition representation encoded
in definition of $X$ and $P$, now $[X_{i} , X_{j}]=0$. This means
that quasiposition \emph{coordinates} are no longer
noncommutative. In fact, noncommutativity at quantum gravity level
now is implemented in the definition of quasiposition
representation and generalization of the notion of spacetime
point.

Now we calculate the scalar product of states $|\phi\rangle$ and
$|\psi\rangle$ in terms of the quasiposition wavefunctions
$\phi(p)$ and $\psi(p)$, namely
$$\langle\phi|\psi\rangle=\int_{-P_{pl}}^{+P_{pl}}\frac{1}{1
-\alpha p + 2\alpha^2 p^2}\phi^*(p)\psi(p)dp=$$
\begin{eqnarray}
{\Lambda_0}^2\int_{-P_{pl}}^{+P_{pl}}\int_{-\infty}^{+\infty}\int_{-\infty}^{+\infty}
\bigg(G_{Mm}(p)\bigg)^{1/2}{e^{\eta\tan^{-1}(\frac{4\alpha p-1}
{\sqrt{7}})}}\times\nonumber\\
e^{i\frac{2(\varepsilon-\varepsilon')}{\alpha\hbar\sqrt
{7}}\bigg(\tan^{-1}(\frac{\eta}{3})+\tan^{-1}(\frac{4\alpha p-1}
{\sqrt{7}})\bigg)}\phi^*(\varepsilon)\psi(\varepsilon')dp\
d\varepsilon\ d\varepsilon'\hspace{5mm}
\end{eqnarray}

where by definition
\begin{equation}
G_{Mm}(p)\ \equiv\ 1 - \alpha p + 2\alpha^2 p^2.
\end{equation}
The difference with original KMM theory comes from the difference
between our $ G_{Mm}(p) $ and that of KMM defined as $ G_{m}(p)\
\equiv\ 1 + \beta p^2$.

\section{ Extension to $n$ dimensions}

\small{}Following the KMM seminal work \cite{1}, in this section
we extend our basic GUP to $n$ dimensions and then we study
modification of the Heisenberg algebra and rotation group due to
this extension.

\subsection{Generalized Heisenberg algebra in $n$ dimensions}

Generalization of the Heisenberg algebra to $n$-dimensions where
rotational symmetry is preserved and there are both a minimal
length and a maximal momentum is
\begin{equation}
[\textbf{X}_i,\textbf{P}_j] = i\hbar\ \delta_{ij} ( 1 - \alpha
\textbf{p} + 2\alpha^2 \vec{\textbf{p}}^2)
\end{equation}
where in three dimensions \ $ \vec{\textbf{p}} = p_x \textbf{i} +
p_y \textbf{j} + p_z \textbf{k} $ \ that \textbf{i}, \textbf{j} and
\textbf{k} are unit vectors of Cartesian coordinates and $
\textbf{p} = \sqrt{\vec{\textbf{p}}^2} $. These commutation
relations imply a nonzero minimal uncertainty in each position
coordinate. As in ordinary quantum mechanics, we have
\begin{equation}
[\textbf{P}_i,\textbf{P}_j] = 0\,.
\end{equation}
So, we can generalize our operators acting on momentum space in
$n$-dimensions as
\begin{equation}
\textbf{P}_i\ \varphi(p) = p_i\ \varphi(p)\,,
\end{equation}
\begin{equation}
\textbf{X}_j\ \varphi(p) = i \hbar\ ( 1 - \alpha \textbf{p} +
2\alpha^2 \vec{\textbf{p}}^2 )\ \frac {\partial}{\partial p_j}\
\varphi(p)\,.
\end{equation}
Then, it is easy to show that
\begin{equation}
[\textbf{X}_i,\textbf{X}_j] = i \alpha \hbar\ (4 \alpha -
\frac{1}{\textbf{p}})\ ( \textbf{P}_i \textbf{X}_j - \textbf{P}_j
\textbf{X}_i )
\end{equation}
Interestingly, now there is a term proportional to
$\frac{1}{\textbf{p}}$ that was absent in original KMM formalism.
This relation reflects the noncommutative nature of the spacetime
manifold in Planck scale. One may worry about divergence for
vanishing momentum. This term originates from introduction of the
$\sqrt{p^2}$ term in our original GUP. A square root is generally
more difficult to handle than polynomials. In our case the
"singularity" arises because the derivative of the square root
diverges at $p=0$. But, fortunately this is not a bad singularity
since the numerator in (75) is linear in $p$ too. We note that in a
more general framework, one should incorporate also the existence of
a minimal measurable momentum. That case is far more difficult than
the current study since both position and momentum space
representation fail to be applicable and one needs to construct a
new and generalized Hilbert space of the model. We are going to
study this issue in a new research program.

Now, if we set $ G_{Mm}(\textbf{p})= 1 - f(\textbf{p}) +
g(\vec{\textbf{p}}^2)$ as a generalization of the previously
defined $G_{Mm}(\textbf{p})$ in Eq.(63), then we find
\begin{equation}
[\textbf{X}_i,\textbf{P}_j] = i\hbar\ \delta_{ij} \Big(\ 1 -
f(\textbf{p}) + g(\vec{\textbf{p}}^2)\Big)\,.
\end{equation}
Then it is straightforward to show that
\begin{equation}
\textbf{X}_j\ \varphi(p) = i \hbar\Big(\ 1 - f(\textbf{p}) +
g(\vec{\textbf{p}}^2)\Big)\ \frac {\partial}{\partial p_j}\
\varphi(p) \,.
\end{equation}
Therefore we find
\begin{equation}
[\textbf{X}_i,\textbf{X}_j] = -i \hbar\Big( \frac{1}{\textbf{p}}
f'(\textbf{p}) + 2 g'(\vec{\textbf{p}}^2)\Big )\Big( \textbf{X}_i
\textbf{P}_j -\textbf{X}_j \textbf{P}_i\Big)
\end{equation}
where by definition
$$f'(\textbf{p})\equiv\frac{df}{d\textbf{p}}\ \
\ \ \ \ \ and\ \ \ \ \ \ \
g'(\vec{\textbf{p}}^2)\equiv\frac{dg}{d\vec{\textbf{p}}^2}$$ For
our case $ f(\textbf{p}) $ and $ g(\vec{\textbf{p}}^2) $ are \ $ -
\alpha \textbf{p} $ \ and \ $ 2\alpha^2 \vec{\textbf{p}}^2 $ \
respectively.

The position and momentum operators are symmetric on the domain of
their definitions with respect to the following scalar product in
$n$ dimensions

\begin{equation}
\langle \Phi | \varphi \rangle\ =\ \int_{-P_{pl}}^{+P_{pl}}\ \frac
{1}{1 - \alpha \textbf{p} + 2\alpha^2 \vec{\textbf{p}}^2}\
\Phi^*(p)\ \varphi(p)\ d^n p\,.
\end{equation}
In this case, the identity operator can be expanded as
\begin{equation}
\textbf{1}\ =\ \int_{-P_{pl}}^{+P_{pl}}\  \frac {1}{1 - \alpha
\textbf{p} + 2\alpha^2 \vec{\textbf{p}}^2}\ |\textbf{p}\rangle
\langle \textbf{p}|\ d^n p\,.
\end{equation}
Therefore, the scalar product of momentum eigenstates in $n$
dimensions is expressed as
\begin{equation}
\langle \textbf{p} | \textbf{p}' \rangle\ =\ ( 1 - \alpha \textbf{p}
+ 2\alpha^2 \vec{\textbf{p}}^2 )\ \delta^n
(\textbf{p}-\textbf{p}')\,.
\end{equation}
At this stage we note that momentum operators can be self-adjoint
in this $n$ dimensional case, but the position operators are
merely symmetric and do not have physical eigenstates (see also
\cite{1}). As what we have done in previous sections for one
dimension, maximal localization states can again be used to define
quasi-position wave functions in $n$ dimensions. The machinery is
the same as we have done for one dimension and we don't repeat it
again. Nevertheless, the quasi-position analysis in $n$
dimensional case is more complicated. Following \cite{1}, now we
focus on rotation group.

\subsection{ The status of the rotation group}

Similar to the KMM scenario, here the rotational symmetry is
respected too. Nevertheless, some modifications are needed due to
the existence of a maximal momentum. Specially, the generalization
to $n$ dimensions proceeds in the same line as KMM theory but now
with a new ingredient originating from maximal momentum. The
generators of rotation in our framework are
\begin{equation}
\textbf{L}_{ij} = \frac {1}{1 - \alpha \textbf{p} + 2\alpha^2
\vec{\textbf{p}}^2}\ ( \textbf{X}_i \textbf{P}_j -\textbf{X}_j
\textbf{P}_i)\,.
\end{equation}
The action on a momentum-space wave function reads
\begin{equation}
\textbf{L}_{ij}\ \varphi(p) = -i \hbar\Big(p_i
\frac{\partial}{\partial p_j} - p_j \frac{\partial}{\partial
p_i}\Big)\ \varphi(p)
\end{equation}
where the following properties are deduced
\begin{equation}
[ \textbf{p}_i,\textbf{L}_{jk} ] = i \hbar\Big( \delta_{ik}
\textbf{p}_j - \delta_{ij} \textbf{p}_k\Big )
\end{equation}

\begin{equation}
[ \textbf{x}_i,\textbf{L}_{jk} ] = i \hbar\Big( \delta_{ik}
\textbf{x}_j - \delta_{ij} \textbf{x}_k \Big)
\end{equation}

\begin{equation}
[ \textbf{L}_{ij},\textbf{L}_{kl} ] = i \hbar\Big( \delta_{ik}
\textbf{L}_{jl} - \delta_{il} \textbf{L}_{jk} + \delta_{jl}
\textbf{L}_{ik} - \delta_{jk} \textbf{L}_{il}\Big)\,,
\end{equation}
and are essentially the same as one encounters in ordinary quantum
mechanics. However, the main change now appears in the relation
\begin{equation}
[\textbf{X}_i,\textbf{X}_j] = -i \alpha \hbar\ (4 \alpha -
\frac{1}{\textbf{p}})\ (1 - \alpha \textbf{p} + 2\alpha^2
\vec{\textbf{p}}^2)\ \textbf{L}_{ij}\,.
\end{equation}
Again, the $\frac{1}{\textbf{p}}$ term which was absent in the
original KMM formalism is a trace of the existence of the maximal
momentum. Once again, this relation reflects the noncommutative
nature of the spacetime manifold in Planck scale.\\

\subsection{ Symmetry and self-adjointness of position and momentum operators}

Finally, the issue of symmetry and self-adjointness of operators
in this setup is important enough to be treated more carefully.
Generally, with a \emph{formally} self-adjoint operator
$\textbf{A}$ in the presence of a minimal measurable length, one
cannot conclude that $\textbf{A}$ is truly a self-adjoint
operator. This is because with a minimal measurable length the
domains $D(\textbf{A})$ and $D(\textbf{A}^{\dag})$ may be so
different in general. By definition, operator $\textbf{A}$ with
dense domain $D(\textbf{A})$ is said to be self-adjoint if
$D(\textbf{A}) = D(\textbf{A}^{\dag})$ and $\textbf{A} =
\textbf{A}^{\dag}$. As KMM have shown, due to existence of a
minimal measurable length, \textbf{X} is a symmetric operator, but
not self-adjoint \cite{1} (see also \cite{37}). In our case, due
to existence of a maximal momentum, the momentum space wave
function $\phi(p)$ vanishes at $p=\pm P_{M}$\, where $P_{M}$ is
the maximal momentum. In this situation, $\textbf{X}$ is a
derivative operator on an interval with Dirichlet boundary
conditions. Nevertheless, $\textbf{X}$ cannot be self-adjoint
since all candidates for the eigenfunctions of $\textbf{X}$ are
not in the domain of $\textbf{X}$ because they obey no longer the
Dirichlet boundary conditions \cite{38}. In fact, the domain of
$\textbf{X}^{\dag}$ is much larger than that of $\textbf{X}$, so
$\textbf{X}$ is indeed not self-adjoint. To be more precise, note
that in our case
$$\int_{-P_M}^{+P_M}\psi^*(p)\bigg(i\hbar\frac{\partial}{\partial
p}\bigg)\phi(p)dp=\int_{-P_M}^{+P_M}\bigg(i\hbar\frac{\partial\psi(p)}{\partial
p}\bigg)^*\phi(p)dp$$
$$\hspace{5.5cm}+i\hbar\psi^*(p)\phi(p)\Bigg|_{-P_M}^{+P_M}\,.$$

Since $\phi(p)$ vanishes at $p=\pm P_M$, then $\psi^*(p)$ can
attain any arbitrary value at the boundaries. The above equation
implies that $\textbf{X}$ is symmetric, but it is not a
self-adjoint operator. In this respect, $\textbf{X}$ acts on
$$D(\textbf{X})=\big\{\phi,\phi'\in L^2\big(-P_M,P_M\big);\phi(P_M)=\phi(-P_M)=0\big\}$$
while $\textbf{X}^{\dag}$ that has the same formal expression,
acts on a different space of functions, namely
$$D(\textbf{X}^{\dag})=\big\{\psi,\psi'\in L^2\big(-P_M,P_M\big)\big\}$$
with no further restriction on $\psi$. Nevertheless, as Kempf has
shown in \cite{39} (see also \cite{1}), there are self-adjoint
extensions of position operator. Since we have worked in the basis
that there is no minimal uncertainty of momentum operator, the
analysis presented in Ref.\cite{39} is essentially applicable to
our case too. In fact, bi-adjoint operator of the densely defined
symmetric position operator is symmetric and closed and has
non-empty deficiency subspaces. From the dimensionalities of these
subspaces one concludes that the position operator is no longer
essentially self-adjoint but has a continuous, one-parameter
family of self-adjoint extensions instead \cite{1}. On the other
hand, the self-adjointness property of $\textbf{P}$ can be proven
by using the von Neumann's theorem (see for instance \cite{40} and
\cite{41}) in the same way as has been shown in \cite{37} and
\cite{38}. We refer the reader to Refs.\cite{42}-\cite{50} for
further developments of these issues.

\section{Summary and Conclusion}

All approaches to quantum gravity proposal support, at least
phenomenologically, the existence of a minimal measurable length
of the order of Planck length. Also, based on Doubly Special
Relativity theories, a test particle's momentum cannot attain any
arbitrary values and is restricted to a maximal value of the order
of Planck momentum. Hilbert space representation of quantum
mechanics with a minimal measurable length has been studied by
Kempf \emph{et al.} \cite{1} (see also \cite{37}, \cite{39} and
\cite{42}-\cite{44}). Here we have generalized the KMM seminal
work to the case that there is also a maximal test particle's
momentum. We have shown that in the presence of both minimal
length and maximal momentum there is no divergency in energy
spectrum of a test particle. Unlike the KMM case that energies of
the short wavelength modes were divergent, in our case there is no
divergency in energy at short wavelengths. As a result, while in
the KMM case, where the quasiposition wavefunctions had no longer
arbitrarily fine ripples, in the presence of maximal momentum
those wavefunctions can have arbitrarily fine ripples. In this
respect unlike the KMM scenario, we obtained correct limiting
equations in the language of the correspondence principle. As we
have shown, position operator $\textbf{X}$ is symmetric but not
self-adjoint in our case. Nevertheless, since there is no minimal
uncertainty in momentum, the self-adjointness of $\textbf{P}$ is
guarantied by the von Neumann's theorem. We note however that even
for the self-adjoint position and momentum operators, it is by no
means obvious that the resulting Hamiltonian for physical systems
will be self-adjoint unless the potential term is specified and
the appropriate domain is chosen. Finally, we note that a more
general treatment of the Hilbert space representation includes
also a nonzero, minimal uncertainty in momentum measurement as
well as position. This general case is far more difficult to
handle since neither a position nor a momentum space
representation is available. This feature can be considered as a
new research program.\\

{\bf Acknowledgement}

We are very grateful to Achim Kempf, Saurya Das, Giovanni
Amelino-Camelia and Pouria Pedram for very fruitful discussions and
suggestions. We also are indebted to two anonymous referees for very
insightful comments.


\begin{thebibliography}{}
\bibitem{1}A. Kempf, G. Mangano and R. B. Mann,   Phys. Rev. D \textbf{ 52}, 1108 (1995).
\bibitem{2}G. Veneziano, Europhys. Lett. \textbf{2}, 199 (1986).
\bibitem{3}D. Amati, M. Cialfaloni and G. Veneziano, Phys. Lett. B \textbf{ 197}, 81 (1987).
\bibitem{4}D. Amati, M. Cialfaloni and G. Veneziano, Phys. Lett. B \textbf{ 216}, 41 (1989).
\bibitem{5}D. J. Gross and P. F. Mende, Phys. Lett. B \textbf{ 197}, 129 (1987).
\bibitem{6}K. Konishi, G. Paffuti and P. Provero, Phys. Lett. B \textbf{ 234}, 276 (1990).
\bibitem{7}R. Guida, K. Konishi and P. Provero, Mod. Phys. Lett. A \textbf{6}, 1487 (1991).
\bibitem{8}M. Kato, Phys. Lett. B \textbf{245}, 43 (1990).
\bibitem{9}L. J. Garay, Int. J. Mod. Phys. A \textbf{10}, 145 (1995).
\bibitem{10}S. Capozziello, G. Lambiase and G. Scarpetta, Int. J. Theor. Phys. \textbf{39}, 15 (2000).
\bibitem{11}M. Maggiore, Phys. Lett. B \textbf{304}, 65 (1993).
\bibitem{12}M. Maggiore, Phys. Lett. B \textbf{319}, 83 (1993).
\bibitem{13}M. Maggiore, Phys. Rev. \textbf{49}, 5182 (1994).
\bibitem{14}F. Scardigli, Phys. Lett. B \textbf{452}, 39 (1999).
\bibitem{15}K. Nozari and T. Azizi, Gen. Rel. Grav. \textbf{38}, 735 (2006).
\bibitem{16}K. Nozari and T. Azizi, Gen. Rel. Grav. \textbf{38}, 735 (2006).
\bibitem{17}K. Nozari and P. Pedram,  Europhys. Lett. \textbf{92}, 50013 (2010).
\bibitem{18}K. Nozari and B. Fazlpour, Chaos, Solitons and Fractals, \textbf{34}, 224 (2007).
\bibitem{19}K. Nozari and S. H. Mehdipour, Gen. Rel. Grav. \textbf{37}, 1995 (2005).
\bibitem{20}K. Nozari, Phys. Lett. B \textbf{629}, 41 (2006).
\bibitem{21}P. Pedram, K. Nozari and S. H. Taheri,  JHEP \textbf{1103}, 093 (2011).
\bibitem{22}P. Pedram, Int. J. Mod. Phys. D \textbf{19}, 2003 (2010).
\bibitem{23}S. Das, E. C. Vagenas, Phys. Rev. Lett. \textbf{101}, 221301 (2008).
\bibitem{24}A. F. Ali, S. Das and E. C. Vagenas,  Phys. Lett. B \textbf{678}, 497 (2009).
\bibitem{25}S. Das and E. C. Vagenas, Can. J. Phys. \textbf{87}, 233 (2009).
\bibitem{26}S. Das, E. C. Vagenas and A. F. Ali, Phys. Lett. B \textbf{690}, 407 (2010).
\bibitem{27}S. Basilakos, S. Das and E. C. Vagenas, JCAP \textbf{09}, 027 (2010).
\bibitem{28}A. F. Ali, S. Das and E. C. Vagenas, Phys. Rev. D {\textbf{84}}, 044013 (2011).
\bibitem{29}R. J. Adler, Am. J. Phys. \textbf{78},925 (2010).
\bibitem{30}S. Hossenfelder,  Class. Quant. Grav. \textbf{25}, 038003
(2008),\\
S. Hossenfelder, [arXiv:1203.6191].
\bibitem{31}P. Wang, H. Yang and X. Zhang, JHEP \textbf{08}, 043 (2010).
\bibitem{32}G. A. Camelia, Int. J. Mod. Phys. D \textbf{11}, 35
(2000); G. Amelino-Camelia, Nature \textbf{418}, 34 (2002); G.
Amelino-Camelia,  Int. J. Mod. Phys. D \textbf{11}, 1643 (2002); J.
Kowalski-Glikman,  Lect. Notes Phys. \textbf{669}, 131 (2005); G.
Amelino-Camelia, J. Kowalski-Glikman, G. Mandanici and A.
Procaccini,  Int. J. Mod. Phys. A \textbf{20}, 6007 (2005); K.
Imilkowska, J. Kowalski-Glikman, Lect. Notes Phys.\textbf{702}, 279
(2006).
\bibitem{33}J. Magueijo and L. Smolin, Phys. Rev. Lett. \textbf{88}, 190403 (2002).
\bibitem{34}J. Magueijo and L. Smolin, Phys. Rev. Lett. D \textbf{ 67}, 044017 (2003).
\bibitem{35}J. Magueijo and L. Smolin, Phys. Rev. D \textbf{71}, 026010 (2005).
\bibitem{36}J. L. Cortes and J. Gamboa, Phys. Rev. D \textbf{71}, 065015 (2005).
\bibitem{37}A. Kempf, Phys. Rev. D \textbf{63}, 024017 (2000).
\bibitem{38}P. Pedram, Phys. Rev. D, \textbf{85}, 024016 (2012).
\bibitem{39} A. Kempf, [arXiv:hep-th/9405067].
\bibitem{40}N. I. Akhiezer and I. M. Glazman, \emph{Theory of Linear
Operators in Hilbert Space}, Dover, New York, 1993.
\bibitem{41}G. Bonneau, J. Faraut and G. Valent, Am. J. Phys. \textbf{69}, 322 (2001).
\bibitem{42}H. Hinrichsen and A. Kempf,  J. Math. Phys. \textbf{37}, 2121 (1996).
\bibitem{43}A. Kempf, J. Math. Phys. \textbf{38}, 1347 (1997).
\bibitem{44}A. Kempf,  Rept. Math. Phys. \textbf{43}, 171 (1999).
\bibitem{45}M. Bojowald and A. Kempf, [arXiv:1112.0994].
\bibitem{46}A. Mostafazadeh, Int. J. Geom. Meth. Mod. Phys. \textbf{7}, 1191 (2010).
\bibitem{47}B. Bagchi and A. Fring, Phys. Lett. A \textbf{373}, 4307 (2009).
\bibitem{48}C. Bender, Rept. Prog. Phys. \textbf{70}, 947 (2007)
\bibitem{49}A. Fring, L. Gouba and F. Scholtz, J. Phys. A: Math. Theor. \textbf{43}, 345401
(2010).
\bibitem{50}A. Fring, L. Gouba and F. Scholtz, J. Phys. A: Math. Theor. \textbf{43},
425202 (2010).

\end{thebibliography}
\end{document}